\documentstyle[seceq,mbf,epsf,wrapft,preprint]{ptptex}

\title{
Excitation Spectrum in the Friedberg-Lee Model}
\author{
Hiroto {\sc Kowata} and 
Masaki {\sc Arima}}

\inst{Department of Physics, Osaka City University, Osaka 558-8585, Japan}

\abst{
 The excitation spectrum of the nucleon with the spin 1/2
is examined by using the Friedberg-Lee 
model containing the constituent quark and the scalar meson.
 An appropriate way of quantization for the non-linear meson field is employed 
by taking account of the non-topological soliton existed in the classical level.
 Our model space for the nucleon resonances includes the three-quark plus 
one-meson state in addition to the pure three-quark state.
 The excitation spectrum in this model space shows that
the positive parity state appears as the first excited state associated 
with the $0s$-excitation of the scalar meson.
 The meson excitation also generates the additional negative parity state 
apart from the well-known $0p$-excitation of the quark.
}
\notypesetlogo

\begin{document}
\maketitle

\section{INTRODUCTION}
 Since the internal structure of hadrons was recognized,
various kinds of quark models were proposed\cite{dgg,ik,glz,bm,cb,cqs}.
 These models have been believed to be successful in describing hadron properties.
 However there are some exceptions for which reasonable explanations have not 
been given yet.
 One of the most difficult things is that the observed first excited state of the 
nucleon has positive parity contrary to many predictions 
commonly obtained by the quark models\cite{pdg}.

 Several works suggested that extra degrees of freedom 
should be introduced so as to reproduce the observed spectrum of 
the nucleon with the spin 1/2,
e.g. surface vibration in the bag model\cite{bdj,g}.
 However plausible candidates for the extra degrees of freedom 
can be found in the quark models;
they usually contain the gluon and/or the meson in addition to the quark.
 The dynamical role of these bosons for the internal structure of baryons
was not investigated sufficiently 
because of the static, and in many cases perturbative, 
treatment applied to\cite{glz,cb}.

 On the other hand, the Skyrme model is another type of the effective model 
which is composed only of the meson field with non-linear self-interaction.
 Although general features of the nucleon resonances are not reproduced,
the positive parity state appears as the first excited state owing to the 
surface vibration of the topological soliton\cite{skyrme}.
 This result indicates that the introduction of quantized meson field with non-linear
self-interaction may solve the parity-ordering problem in the quark models.

 In order to realize this idea, the soliton bag model originally 
proposed by Friedberg and Lee is one of the noticeable models
which treat the meson field non-perturbatively \cite{FL_o,lp}.
 The non-topological soliton of the scalar meson is formed in this model 
due to the non-linear self-interaction for the meson and
its coupling with the quarks.
 This soliton behaves as an effective binding potential for the quarks.

 It is interesting to study the nucleon resonances in this model
for exploring the dynamical role of the mesons.
 In the existing works dealing with the nucleon resonances 
in the Friedberg-Lee model\cite{haider,sskh},
the mean-field approximation was applied to the scalar meson field 
instead of properly quantizing it.
 Thus the excitation spectrum obtained therein exhibited the same character 
as usual: the first excited state was predicted to have negative parity.
 Furthermore the orthogonality among the quark wave functions belonging 
to different excited states was lost in their treatment.

 In this paper, we investigate the role of the scalar meson in the 
structure of the nucleon resonances with the spin 1/2
by using the Friedberg-Lee model.
 Although this model does not include the pion field
which is also important in low-energy hadron phenomenology,
static properties of the nucleon can be reproduced 
by this model reasonably well\cite{haider-liu}.
 We take this simple model as suitable one for the first step of our exploration.
 We quantize the scalar meson field by introducing the fluctuation 
around the non-topological soliton\cite{gw,Jackiw,rajaraman}.
 Note that the scalar meson field in the Friedberg-Lee model does not 
necessarily correspond to the so-called $\sigma$-meson as a chiral partner 
of the $\pi$-meson (or as a $\pi\pi$ $s$-wave resonance).
 In this work, according to the original idea mentioned in Ref.~\citen{FL_o},
we consider this scalar meson field as a purely phenomenological one
to generate the binding potential for the quark.
 Examination of the properties of this scalar meson is addressed 
to our future work in which our model will be improved 
by taking into account the chiral symmetry.

 This paper is organized as follows.
 In section 2, we give the formulation of our model.
 For simplicity, we replace the relativistic quark
by the non-relativistic one.
 Calculating a spectrum for several low-lying states of the nucleon resonances, 
we examine whether the Friedberg-Lee model with the quantized scalar meson
properly predicts the observed sequence of intrinsic parity.
 Our results and discussion are given in section 3.

\section{FORMALISM}
\label{sec:formalism}

 The Friedberg-Lee model with the non-relativistic quark field is based on the Hamiltonian,
\begin{eqnarray}
H &=& \int d^3r \left[\psi^{\dagger}\left(-\frac{\nabla^2}{2m} + g\sigma\right)\psi\right.
+ \left.\frac{1}{2}\Pi^2 + \frac{1}{2}(\nabla\sigma)^2 + U(\sigma)\right],
\label{eqn:h_o}
\end{eqnarray}
where $\psi$ is the two-component spinor field for the constituent quark.
 The scalar meson field and its conjugate field are denoted 
by $\sigma$ and $\Pi$, respectively.
 The potential $U(\sigma)$ is explicitly given by
$a\sigma^2(\sigma-\sigma_v)^2/2$, which involves non-linear terms
up to fourth order of $\sigma$.
 We do not employ more general form of $U(\sigma)$ in order to reduce 
the number of phenomenological parameters.
 The parameters, $a$, $\sigma_v$, $g$ and $m$, stand for
the potential strength, the vacuum expectation value of $\sigma$,
the meson-quark coupling constant 
and the constituent quark mass, respectively.
 The Hamiltonian (\ref{eqn:h_o}) does not include direct interactions
among the quarks such as the one-gluon-exchange potential.

 Here we quantize the quark and the scalar meson fields.
 The conventional canonical method is applied to the quark field:
\begin{equation}
\psi({\mbf r},t) = \sum_n c_ne^{-i\epsilon_nt}\psi_n({\mbf r}),
\label{eqn:ex-quark}
\end{equation}
where the annihilation and creation operators, $c_n$ and $c_n^{\dagger}$,
satisfy the usual anti-commutation relation.
 In the case of the scalar meson field, which is assumed to form the non-topological
soliton in the classical treatment,
we quantize the fluctuation around this non-topological soliton in 
the canonical method;
\begin{eqnarray}
\sigma({\mbf r},t) &=& \sigma_0({\mbf r}) + \chi({\mbf r},t),
\nonumber\\
 &=& \sigma_0({\mbf r}) +\sum_k\frac{1}{\sqrt{2\omega_k}}
[b_ke^{-i\omega_kt}s_k({\mbf r}) + b_k^{\dagger}e^{i\omega_kt}s_k^*({\mbf r})],
\label{eqn:ex-meson}
\end{eqnarray}
where the annihilation and creation operators, $b_k$ and $b_k^{\dagger}$, 
satisfy the usual commutation relation.
 Note that the meson field is not merely 
the semiclassical background for the constituent quark.

 The complete orthogonal sets, $\psi_n({\mbf r})$ and $s_k({\mbf r})$,
and the eigenenergies, $\epsilon_n$ and $\omega_k$, in the expansions 
(\ref{eqn:ex-quark}) and (\ref{eqn:ex-meson}) are specified by 
solving the eigenvalue equations,
\begin{eqnarray}
\left(-\frac{\nabla^2}{2m} + g\sigma_0({\mbf r})\right)\psi_n({\mbf r})
&=&\epsilon_n\psi_n({\mbf r}),
\label{eqn:quark-sp}
\\
\left(-\nabla^2 + U^{\prime\prime}(\sigma_0({\mbf r})) 
- \omega^2_k\right)s_k({\mbf r}) &=& 0,
\label{eqn:meson-sp}
\end{eqnarray}
where $n$ and $k$ abbreviate the quantum numbers of 
the quark and the scalar meson, respectively.
 On deriving Eq.(\ref{eqn:meson-sp}), we assume that the perturbative 
treatment is possible for the field $\chi$.
 And hence we limit the number of the scalar meson to one and zero.

 The non-topological soliton solution $\sigma_0({\mbf r})$ appeared 
in Eqs.(\ref{eqn:ex-meson}),
(\ref{eqn:quark-sp}) and (\ref{eqn:meson-sp}) is obtained by solving
\begin{eqnarray}
\nabla^2\sigma_0({\mbf r}) - U^{\prime}(\sigma_0({\mbf r}))
&=& 3g\psi_0^{\dagger}({\mbf r})\psi_0({\mbf r}),
\label{eqn:soliton}
\end{eqnarray}
where $\psi_0({\mbf r})$ is the lowest-energy solution $(n=0)$ of Eq.(\ref{eqn:quark-sp}).
 When we specify the right hand side of Eq.(\ref{eqn:soliton}),
all three quarks are assumed to be in the ground state.
 To determine $\sigma_0({\mbf r})$ and $\psi_0({\mbf r})$,
we solve the coupled equations, Eq.(\ref{eqn:quark-sp}) for $n=0$ and Eq.(\ref{eqn:soliton}).
 The non-linear terms included in $U^{\prime}(\sigma)$
and the coupling between the quark and the meson lead to the existence 
of the soliton solution $\sigma_0({\mbf r})$.
 Then we obtain $\psi_n({\mbf r})$ for other $n$ and $s_k({\mbf r})$
by solving Eqs.(\ref{eqn:quark-sp}) and (\ref{eqn:meson-sp}).

 The soliton solution also behaves as the effective binding potential for the quarks,
which has the asymptotic value $g\sigma_v$ for infinitely large value of $|{\mbf r}|$.
 Therefore, in addition to the bound state solution, there exist continuum states.
 In the following calculations, we do not take into account several quantum corrections
such as a loop correction.
 We consider only the three-quark bound states for the nucleon and neglect 
all contributions of the continuum states.

 Here we construct the single baryon states in the Fock space defined by the operators
in Eqs.(\ref{eqn:ex-quark}) and (\ref{eqn:ex-meson}).
 They are classified into two types:
the pure three-quark state and the three-quark plus one scalar meson state.
 We use the ket $|N,LS;J\rangle$ for the pure three-quark state 
where $N$ means the collection of the spatial 
quantum number $(n)$ and the spatial symmetry,
$L$ is the total orbital angular momentum, $S$ is the intrinsic spin 
and $J$ is the total spin of the state.
 For example, $N=[(0s)(0p)^2]_{M}$ means that two quarks 
excite to the $0p$-orbit while the other one is in the $0s$-orbit.
 This state has the mixed symmetry with respect to the particle exchange
denoted by the subscript $M$.
 We take a symmetric combination of the spatial and the SU(6) spin-flavor 
quantum numbers for the three-quark state
because the color part is always antisymmetric by itself.
 We write the three-quark plus one meson state as 
\begin{equation}
|N,LS;J_{3q},l;J\rangle = \left[b_k^{\dagger}|N,LS;J_{3q}\rangle\right]_J
\end{equation}
where $J_{3q}$ is the total spin of the three-quark state,
$b_k^{\dagger}$ operates on the meson vacuum,
that is the pure three-quark state.
 Inclusion of this state distinguishes our study from 
the conventional ones because they usually employ only the pure three-quark state
to describe baryon resonances.

 By using the expansions (\ref{eqn:ex-quark}) and (\ref{eqn:ex-meson}),
the Hamiltonian (\ref{eqn:h_o}) becomes
\begin{eqnarray}
H &=& H_0 + H_I,
\label{eqn:hamiltonian}
\\
H_0 &=& E_{\sigma} + \sum_{n}\epsilon_nc_n^{\dagger}c_n
+ \sum_k\omega_kb_k^{\dagger}b_k,
\label{eqn:h0}
\\
H_I &=& \int d^3r \left(~g\psi^{\dagger}\psi\chi
-g\langle 0|\psi^{\dagger}\psi|0\rangle\chi + O(\chi^3, \chi^4)~\right),
\end{eqnarray}
where $E_{\sigma}$ is the classical energy of the non-topological soliton
and $|0\rangle$ indicates all three quarks occupy the ground state with no meson.
 The difference between the present model and the usual Friedberg-Lee model
is the third term in Eq.(\ref{eqn:h0}) and $H_I$,
which appear since we quantize the scalar meson field.
 In our model the scalar meson contributes to the excitation energies 
if the baryon state includes one meson.
 There exists explicit coupling of the quark with the meson in $H_I$.
 Because of the perturbative treatment of $\chi$, 
higher order corrections due to $O(\chi^3, \chi^4)$ terms are not considered here.

 To obtain the energy eigenvalues of the Hamiltonian (\ref{eqn:hamiltonian}),
we take the diagonalization method.
 As for the basis states, we use the single baryon states given above.

\section{RESULTS and DISCUSSION}
\label{sec:result}

\subsection{parameters}
\label{subsec:prm}
 We set the constituent quark mass at 300 MeV as usually used in 
various non-relativistic models.
 As for three other parameters, $a$, $\sigma_v$ and $g$,
we determine their values by referring to the observed spectrum for 
the spin 1/2 nucleon resonances:
the first and second excited states are the positive and negative parity states,
and their excitation energies are about 500 and 600 MeV, respectively.
 Many results of the constituent quark models suggest that the mass of this positive parity 
state cannot be explained by the orbital or nodal excitations of the quark
while the negative parity state is described by the orbital excitation 
of the quark\cite{neg-p}.
 This observation motivates us to find parameter sets 
which satisfy the following condition:
the first excited state is due to the one meson $0s$-excitation,
instead of the quark excitation, with the excitation energy of 500 MeV
and the second one to the orbital $0p$-excitation of the quark
with the excitation energy of 600 MeV.

\begin{figure}[htb]
 \parbox{\halftext}{
  \epsfxsize=6.6cm
  \centerline{\epsfbox{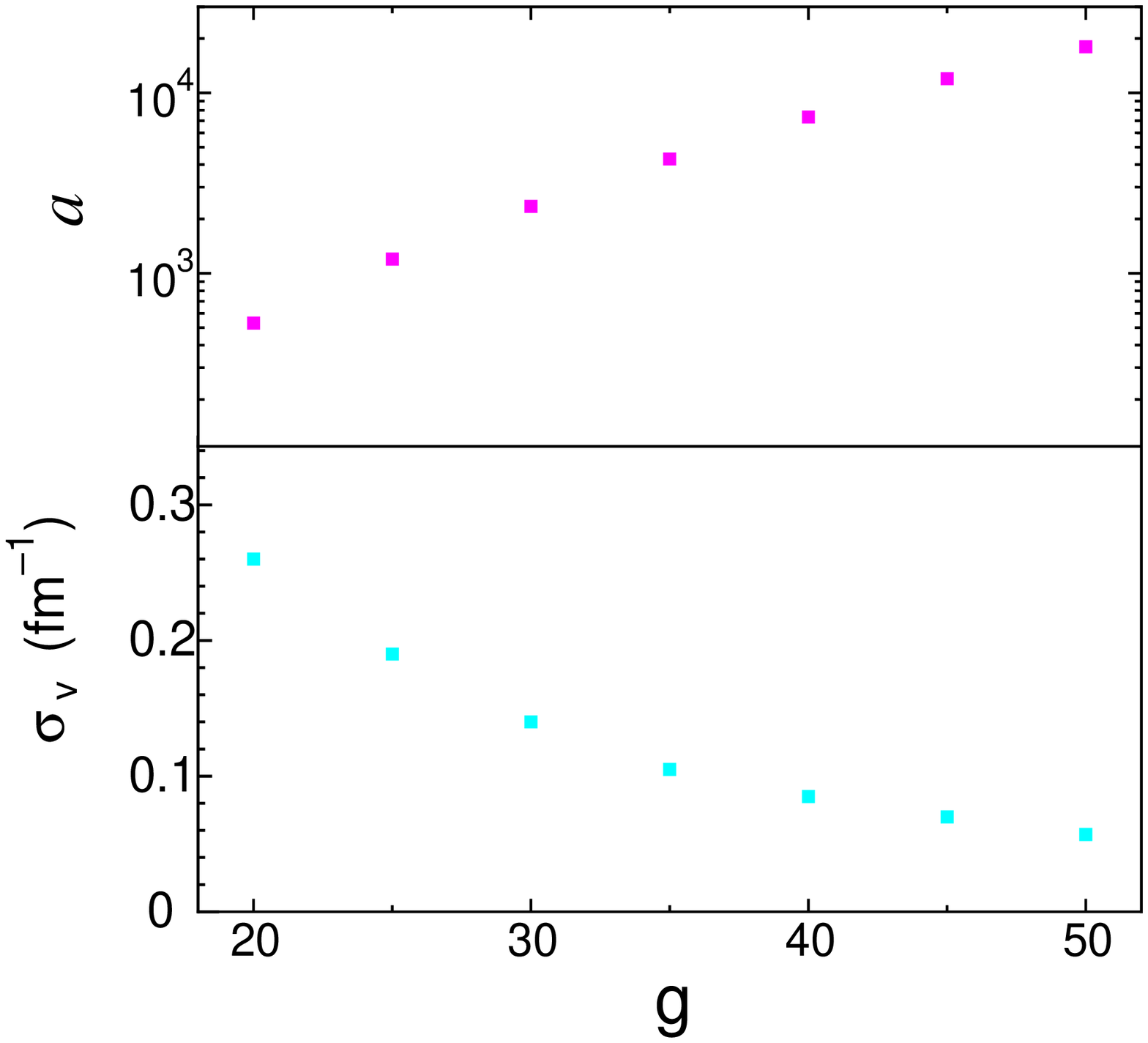}}
  \caption{
  The variation of the parameters $a$ and $\sigma_v$ as functions
  of the meson-quark coupling constant $g$.}
 \label{fig:pd1}}
 \parbox{\halftext}{
  \epsfxsize=6.6cm
  \centerline{\epsfbox{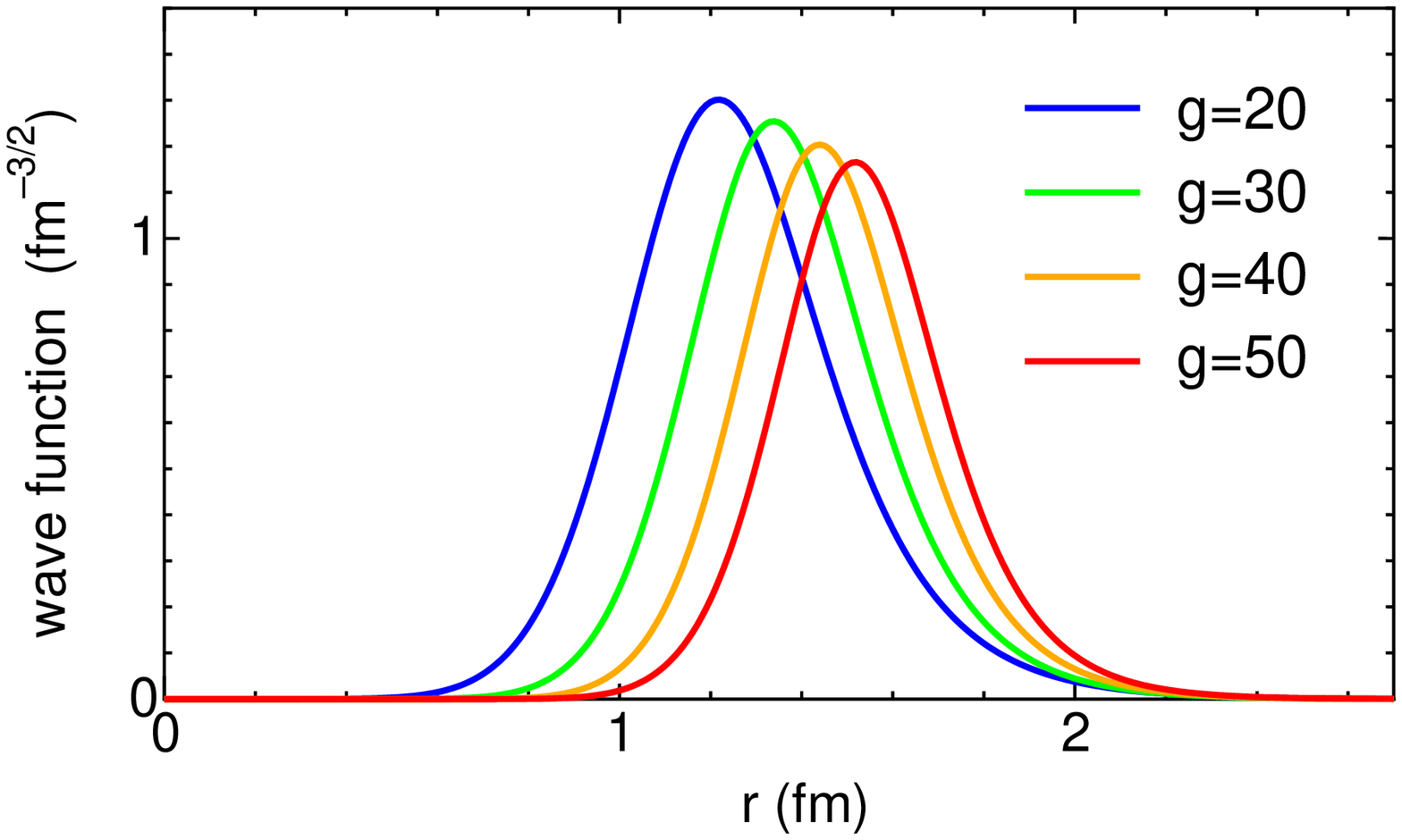}}
  \caption{
  The $0s$-meson wave functions for each parameter set.}
\label{fig:pd2}}
\end{figure}

 The results of our parameter search for $a$ and $\sigma_v$
are shown in Figure \ref{fig:pd1} as functions of $g$.
 We find monotonic behavior of these parameters.
 For the smaller value of $g$ than 20, the quark single-particle state
fails to have bound states above the $0p$-orbit.
 In the following calculation we consider only those parameters
by which there are at least four bound states, $0s$, $0p$, $0d$ and $1s$-orbits
for the quark single-particle state.
 For larger value of $g$ than 50, although we do not draw curves in Figure \ref{fig:pd1},
we can easily obtain the values of $a$ and $\sigma_v$ by using extrapolation.

 Now let us calculate other quantities obtained by the single-particle wave functions
and examine their dependence on the parameters given in Figure \ref{fig:pd1}.
 First we calculate the root mean square radius of 
the $0s$-quark wave function.
 We find that this quantity is nearly constant at about 0.56 fm for each value of $g$
because the potential felt by the $0s$-quark does not change so much.

 However we can see difference of each parameter set 
if we observe the $0s$-meson wave function.
 In Figure \ref{fig:pd2}, we find out that
the $0s$-meson wave function is gradually pushed out for larger value of $g$.
 The wave functions of other excited states show
the same behavior as the $0s$-meson wave function.
 Thus we can specify the best-fit parameters
if there are some observables which are sensitive to the
change in the root mean square radius of the meson wave function.
 But unfortunately we have no information about such observables.

\subsection{spurious states}
 Before going into the diagonalization, we consider the spurious states inevitably 
appeared in our formalism.
 These states do not describe the internal excitation but the excitation of 
the center of mass (c.m.) motion.
 To remove them from our basis states, the shell-model technique
is useful for our practical calculation: we project out the spurious states
by calculating the matrix element of an artificial potential $V({\mbf R})$ 
which depends only on the c.m. coordinate ${\mbf R}$ of the system\cite{dsf}.
 We take the harmonic oscillator form for $V({\mbf R})(\propto{\mbf R}^2)$ 
to simplify the calculation.

 We take special notice of the spurious states appearing in negative parity states 
since these states will correspond to
the negative parity resonances observed confidently in experiments.
 First we consider the pure three-quark states with the 0$p$-excitation of one quark.
 Applying the above technique, we can extract two states out of three states 
belonging to these negative parity states:
$\left|[(0s)^2(0p)]_M,1s;\frac{1}{2}\right\rangle$, $(s=\frac{1}{2},\frac{3}{2})$.
 Next we consider the three-quark plus one meson states when one of their constituents 
excites to the 0$p$-orbit.
 We obtain three states after removing the spurious state:
$\left|[(0s)^2(0p)]_M,1s;\frac{1}{2},0;\frac{1}{2}\right\rangle$, 
$(s=\frac{1}{2},\frac{3}{2})$
and 
$\lambda_1\left|[(0s)^3]_S,0\frac{1}{2};\frac{1}{2},1;\frac{1}{2}\right\rangle
+\lambda_2\left|[(0s)^2(0p)]_S,1\frac{1}{2};\frac{1}{2},0;\frac{1}{2}\right\rangle$
where $\lambda_1\approx\lambda_2$ which is found by the numerical calculation.
 On the other hand, the contributions from the spurious energy which is 
inevitably included in this system are not removed in this prescription.
 Although a few methods are proposed, for example Ref.~\citen{lbhw},
no reliable one is established until now in order to resolve this problem.
 Therefore we do not touch on this problem but leave as a future subject.
 
 There is no effect of the spurious motion 
on the first positive parity excited state.
 As for other positive parity states, 
because they appear in the higher energy region,
we do not touch on the spurious states here.

\subsection{excitation spectra}
\begin{figure}
\epsfxsize=10cm
 \centerline{\epsfbox{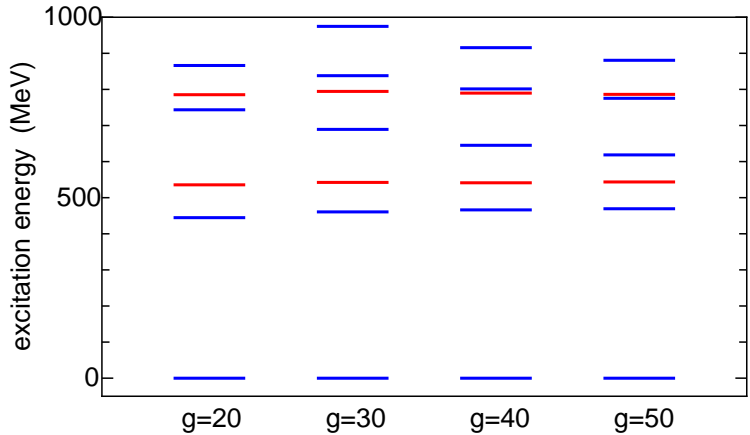}}
 \caption{
 The excitation spectra of the spin 1/2 nucleon resonance
 for the four parameter sets.
 The positive and negative parity states are indicated by 
 the blue and the red lines, respectively.
 The energy of the ground state is set to zero.
 The negative parity states above the first excited state are doubly degenerate.}
 \label{fig:sp}
\end{figure}

 We consider the excitation spectra obtained by diagonalizing the 
 Hamiltonian using several parameter sets obtained above.
 The number of the basis states is 14 for the positive parity states,
and 8 for the negative parity states.
 We show several lower basis states in Table \ref{tbl:bs}
and the excitation spectrum in Figure \ref{fig:sp}.
 The ground state energy is set to zero in this figure.
 Our effective model is developed so as to reproduce the observed structure 
of the nucleon, but this model is not matured enough to determine the absolute 
value of the ground state energy.

\begin{table}[h]
\begin{tabular}{c|c}
\hline\hline
positive parity & negative parity \\
\hline
\\
$\begin{array}{l}
\left|[(0s)^3]_S,0,\frac{1}{2},\frac{1}{2}\right\rangle\\\\
\left|[(0s)^3]_S,0,\frac{1}{2};\frac{1}{2},0;\frac{1}{2}\right\rangle
\end{array}$ &
$\begin{array}{ll}
\left|[(0s)^2(0p)]_S,1,\frac{3}{2},\frac{1}{2}\right\rangle
& \left|[(0s)^2(0p)]_M,1,s,\frac{1}{2}\right\rangle, (s=\frac{1}{2}, \frac{3}{2})\\\\
\left|[(0s)^3]_S,\frac{1}{2};\frac{1}{2},1;\frac{1}{2}\right\rangle
& \left|[(0s)^2(0p)]_M,0,s;\frac{1}{2},0;\frac{1}{2}\right\rangle, 
(s=\frac{1}{2}, \frac{3}{2})\\\\
\left|[(0s)^2(0p)]_S,0,\frac{3}{2};\frac{1}{2},0;\frac{1}{2}\right\rangle &
\end{array}$ \\\\
\hline
\end{tabular}
\caption{
Several examples of the basis states.
The spatial symmetry ($S$ or $M$) is indicated by 
the irreducible representation of $S_3$.
}
\label{tbl:bs}
\end{table}

 The following features are common for all spectra obtained for each value of $g$.
 The ground state is mainly composed of the three-quark state in which
all three quarks occupy the $0s$-orbit.
 This is consistent with many results of other quark models.

 We obtain the positive parity state as the first excited state.
 Its main component is the three $0s$-quark plus one $0s$-meson.
 This result is established because we can find out the appropriate 
parameter sets satisfying the condition given in \ref{subsec:prm}.
 We also find that the meson-quark interaction $H_I$ has only weak influence 
on the structure of each state because of small overlap 
between the quark and the meson wave functions:
the quark wave function is concentrated in the central region 
of the nucleon by $g\sigma_0({\mbf r})$
while the meson wave function is distributed around the surface 
by $U^{\prime\prime}(\sigma({\mbf r}))$.

 Brown {\it et al.} introduced the breathing mode in the MIT bag model 
to explain the excitation energy of the Roper resonance.
 The distribution of the quantized bag surface is very similar to 
that of the scalar meson in our model.
 And it is worth noting that both the bag and the scalar field are related to
the confinement phenomenon.
 Thus our model can also be interpreted as 
microscopic expression of the breathing mode.

 However we stress here that we can reproduce the parity of the first excited state
without introducing any artificial degrees of freedom.
 The surface of the MIT bag model has no dynamical origin, 
and the quantization was carried out in a somewhat {\it ad hoc} way.
 On the other hand, the scalar meson field is one of the dynamical 
degrees of freedom in our model from the start,
and is treated on an equal footing with the quarks.

 The first excited state is due to the $0s$-excitation of the scalar meson
in the nucleon, or equivalently the monopole mode of the fluctuation 
of the soliton surface.
 This structure is consistent with the idea of Morsch {\it et al.}\cite{morsch}.
 They consider the Roper resonance as the monopole excitation of 
the nucleon (the compression mode)
by the analysis of the scalar-isoscalar excitation of the nucleon 
due to $\alpha$-$p$ scattering.

 Now we turn to the negative parity states.
 Above the first excited state, there are two negative parity states ($S=1/2$ and $3/2$)
degenerate in energy.
 Their main components are the three-quark state with the $0p$-excitation of one quark.
 Their internal structure is consistent 
with many results of other quark models\cite{neg-p},
and they are often allocated to the observed resonances $N(1535)$ and $N(1650)$.
 The reason for this mass difference is usually attributed to
the tensor interaction of the one-gluon-exchange or one-meson-exchange potentials
which are not included in our present model.

 In addition to these two states, another 
negative parity state appears above them by about 250 MeV in our model,
which is mainly composed of the three-quark plus one meson state.
 The appearance of this state is characteristic of our model.
 The correspondence of this state with experimental observation is not known
because there are no observed states in neighborhood of 
well-known negative parity states.
 But we consider this state is neither superfluous nor fatal disease in our model.
 Before allocating our excited states to the observed resonances,
we must apply the present model to the pion-nucleon reaction.
 It is quite possible that the pion-probe can not see this third negative 
parity state due to its weak coupling with the pion-nucleon channel.

\begin{figure}
\epsfxsize=10cm
\centerline{\epsfbox{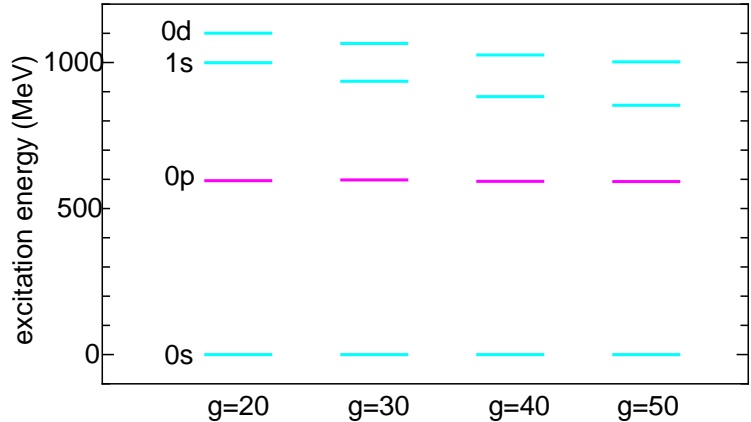}}
\caption{
The quark single-particle energy levels for each parameter set.
The energy difference between $0s$ and $0p$ is constrained to be 600 MeV.
}
\label{fig:pd3}
\end{figure}

 Finally we consider the $g$-dependence of the spectrum.
 The first and second excited states show merely weak $g$-dependence because 
the excitation energies of their main components,
the $0s$- and $0p$-quarks and the $0s$-meson, 
are constrained in the parameter searching.
 On the contrary to these states, other excited states show prominent $g$-dependence.
 The positive parity excited states shift downward as $g$ increases, 
while the negative parity states remain at nearly same energies.
 This behavior is not due to the quark-meson interaction $H_I$, but is closely 
related with the $g$-dependence of the quark single-particle energy level.
 The positive parity excited states considered here include the 
quark in the $1s$- or $0d$-orbit.
 These single-particle states shift downward as $g$ grows as shown in Figure \ref{fig:pd3}.
 On the other hand, the negative parity states include the quark in the 
$0p$-orbit whose excitation energy is constrained 
by the condition given in \ref{subsec:prm}.

 In case of large value of $g$, 
we find the positive parity states with relatively small excitation energies
which may not correspond to the actual nucleon resonance
even if we project out the spurious states.
 From this observation, large value of $g$ may not be appropriate.
 Furthermore, the work of Ref.~\citen{haider-liu} 
indicates the use of small values for $g$ around 10.

\section{SUMMARY}
 We have calculated the excitation spectrum of nucleon resonances with the spin 1/2 
in the Friedberg-Lee model.
 We have quantized the scalar meson field around the classical solution for
the non-topological soliton instead of using the mean-field approximation.
 The positive parity state appears as the first excited state in our calculation,
which is consistent with the observed nucleon resonance spectrum.
 The first excited state is mainly composed of 
the three $0s$-quark plus one $0s$-meson,
and we may correspond this first excited state to the Roper resonance.
 This interpretation is significantly different from the conventional one
in the quark models:
for example, in the constituent quark models with 
the harmonic oscillator confinement potential,
the Roper resonance has been described as 
a $2\hbar\omega$ excitation of the constituent quark.
 Above this state, there appear two degenerate negative parity states.
 Their structures are consistent with the usual quark models.
 There is another negative parity state above these two states,
which is characteristic of the present model.

 We have shown that the dynamical treatment of the scalar meson
in the constituent quark model
may solve the parity-ordering problems in the nucleon spectrum.
 Thus the Friedberg-Lee model with quantized scalar meson field is worth studying further
as an effective model for baryon resonances.

 To improve our present model, we must include the pion field 
which is important in the low energy phenomenology.
 This can be achieved along the line of Ref.~\citen{cs}.
 And we will also apply our model to the analysis of the pion-nucleon reactions.
 This application is important to confirm 
the idea that the nucleon resonances are described 
as composites of the quarks and the mesons.

\section*{Acknowledgement}
The authors thank Prof.\ T. Sato for useful comments.
We also thank Prof. K.\ Masutani and 
Prof. Y.\ Sakuragi for useful discussions.

\end{document}